\documentclass[a4paper,fleqn]{cas-dc}

\usepackage[numbers]{natbib}
\usepackage{amsmath}
\usepackage{xcolor}
\usepackage{graphicx}
\usepackage{float}
\usepackage{multirow}

\def\tsc#1{\csdef{#1}{\textsc{\lowercase{#1}}\xspace}}
\tsc{WGM}
\tsc{QE}

\begin{document}
\let\WriteBookmarks\relax
\def\floatpagepagefraction{1}
\def\textpagefraction{.001}

\shorttitle{Localization of the Ventricular Excitation Origin Without Patient-Specific Geometries}
\shortauthors{Pilia, Schuler, Rees et al.}  

\title[mode=title]{Non-invasive Localization of the Ventricular Excitation Origin Without Patient-specific Geometries Using Deep Learning}

\author[1]{Nicolas Pilia}[orcid=0000-0001-8247-379X]
\cormark[1]
\ead{publications@ibt.kit.edu}
\credit{Supervision, Methodology, Software, Conceptualization, Investigation, Validation, Visualization,  Writing - original draft, Writing - review \& editing}

\author[1]{Steffen Schuler}[orcid=0000-0001-6515-3807]
\cormark[1]
\credit{Supervision, Methodology, Software, Conceptualization, Investigation, Validation, Visualization, Writing - original draft, Writing - review \& editing}

\author[1]{Maike Rees}[orcid=0000-0002-7674-9144]
\cormark[1]
\credit{Methodology, Software, Conceptualization, Investigation, Validation, Writing - review \& editing}

\author[1]{Gerald Moik}[orcid=]
\credit{Methodology, Software, Conceptualization, Investigation}

\author[2]{Danila Potyagaylo}[orcid=]
\credit{Conceptualization, Investigation, Clinical Data Generation}

\author[1]{Olaf Dössel}[orcid=]
\credit{Funding acquisition, Project administration, Supervision, Conceptualization, Writing - review \& editing}

\author[1]{Axel Loewe}[orcid=0000-0002-2487-4744]
\credit{Funding acquisition, Project administration, Supervision, Conceptualization, Writing - review \& editing}

\affiliation[1]{
	organization={Institute of Biomedical Engineering, Karlsruhe Institute of Technology (KIT)},
	city={Karlsruhe},
	country={Germany}
}

\affiliation[2]{
	organization={EPIQure GmbH},
	city={Karlsruhe},
	country={Germany}
}

\cortext[1]{These authors contributed equally.}

\begin{abstract}
Cardiovascular diseases account for 17 million deaths per year worldwide. Of these, 25\% are categorized as sudden cardiac death, which can be related to ventricular tachycardia (VT). This kind of arrhythmia can be caused by focal activation sources outside the sinus node. A curative treatment is catheter ablation of these foci in order to inactivate the abnormal triggering activity. However, the localization procedure is usually time-consuming and requires an invasive procedure in the catheter lab. To facilitate and expedite the treatment, we present two novel localization support techniques based on convolutional neural networks (CNNs) addressing these clinical needs. In contrast to existing methods, our approaches were designed to be independent of the patient-specific geometry and directly applicable to surface ECG signals, while also delivering a binary transmural position. Furthermore, one of the method's outputs could be interpreted as several ranked solutions. The CNNs were trained on a data set containing only simulated data and evaluated both on simulated and clinical test data. On simulated data, the median test error was below 3\,mm. The median localization error on the unseen clinical (test) data was between 32\,mm and 41\,mm without optimizing the pre-processing and CNN on the clinical (test) data. Interpreting the output of one of the approaches as ranked solutions, the best median error of the top-3 solutions dropped to 20\,mm on the clinical test set. The transmural position was correctly detected in up to 82\% of all clinical cases. These results demonstrate a proof of principle to utilize CNNs to localize the activation source without the intrinsic need of patient-specific geometrical information. Furthermore, delivering multiple solutions can help the physician to find the real activation source amongst more than one possible locations. With a further optimization to clinical data, these methods have a high potential to speed up clinical interventions, replace certain steps within those and therefore consequently decrease procedural risk and improve VT patients' outcomes.
\end{abstract}




\maketitle

\section{Introduction}
\label{intro}
Sudden cardiac death (SCD) is a relevant cardiovascular disease. Approximately 4.25 million deaths per year worldwide are related to SCD~\cite{Priori-2015-ID16635}, often due to ventricular tachycardia (VT). VT can be caused by structural heart diseases but is also found in patients with structurally healthy hearts. The electrocardiographic manifestations of VT  evaluated in the 12-lead electrocardiogram (ECG) are commonly characterized by a changed QRS complex~\cite{Willems-1985-ID16634,Griffith-1994-ID16633}. The so-called monomorphic VT presents a repetitive sequence of equally shaped QRS complexes as the underlying ventricular activation sequence remains the same from beat to beat. This hints either at substrate supporting a stable re-entrant mechanism or at a single focus region triggering the activation~\cite{Roberts-Thomson-2011-ID16632}. The standard treatment for this kind of VT includes medication but also radio-frequency ablation, where the triggering region is prevented to sustain the arrhythmia by localized thermal destruction~\cite{Roberts-Thomson-2011-ID16632}. However, these regions need to be localized precisely in order to destroy the perturbing region and spare healthy tissue. Activation mapping and pace mapping are the two standard procedures to achieve this. In both procedures, catheters need to be inserted into the ventricle. By analyzing measured intracardiac signals, the triggering site can often be identified. These kind of procedures are minimally invasive but still suffer from drawbacks such as their long duration and the risks e.g. for heart perforation~\cite{JOSHI-2005-ID16631}. 

To speed up the localization of the triggering region and minimize the risk for the patient, several techniques have been proposed. The first is the classification by patterns visible in the single leads of the 12-lead surface ECG~\cite{Roberts-Thomson-2011-ID16632,Segal-2007-ID13248, Anderson-2019-ID12751}. This is relatively easy to apply but requires experienced cardiologists and works only for the locations covered by one of the rules. 

As an alternative, automatic computerized approaches exist that can directly predict the location of the ablation targets. One of these is utilizing ECG imaging (ECGI) techniques where an electrode grid is placed on the torso of the patient, often consisting of more than 30 electrodes, to record the so-called body surface potential map (BSPM). The BSPM is then used to solve the inverse problem of ECG by calculating the electrical activity on the heart surface, for example described by transmembrane voltage courses. Knowing these signal courses at different locations, the point in space with earliest activation can be determined. However, solving the inverse problem normally requires a prerecorded patient-specific geometry, e.g. from magnetic resonance imaging (MRI) and is sensitive to the used methods, the pre-processing of patient imaging data and the parameterization of both steps~\cite{Rudy_2017,Tate_2018,Bear_2018}. 

These drawbacks are one main motivation for applying machine learning to the problem of locating the triggering focal activation source, i.e. the origin of the so-called ventricular ectopic beat. Several studies already proved the applicability of machine learning approaches to the problem of localizing the activation wave source with simulated signals, partly by first estimating the activation times on the heart surface, partly by using pacing signals~\cite{unknown-0000-ID16466,Bacoyannis-2021-ID15881}. Learning from patient-specific simulated signals is another already used approach which helps to estimate activation times and the location of the activation source on clinical data~\cite{Giffard-Roisin-2019-ID13213,Alawad-2019-ID12834}. Besides, simulated data can help to overcome the shortage of clinical data in this field, which often prevents the use of machine learning approaches. Instead of directly estimating a set of coordinates as the origin of the activation, it is also possible to divide the ventricles into segments and predict each segment's probability to contain the activation wave source~\cite{Kaiyue-2020-ID13709,unknown-0000-ID16466,Zhao_2022}. The latter aims at overcoming the problem of inter-patient variability as these segments can be defined on all ventricles. An alternative to the introduction of ventricular segments are specifically designed machine learning approaches~\cite{Gyawali-2020-ID14304} or the use of a ventricular coordinate system~\cite{unknown-0000-ID16466}. Fully automatic approaches still relying on patient-specific geometries or usable for only one ventricle were successfully implemented in~\cite{Sapp-2017-ID15770,Zhou-2019-ID13157,Yang-2018-ID12792}.
A very recent work classified the presence of ventricular outflow tract arrhythmia fully automatically using simulations to boost results on patient data ~\cite{Doste_2022}. 

All the methods mentioned above either lack the general applicability to clinical data without the need for the patient-specific geometry (i.e., additional imaging procedures), are only tested on simulated data (i.e., did not yet prove to overcome the domain gap), or are constrained to only a part of the ventricles or need to be tuned for every patient (i.e., are not universally applicable and not generalizing).

In the following, we present a complete pipeline to localize the source of activation directly from a BSPM with several CNNs overcoming the mentioned drawbacks of existing methods. With our methods it is possible to determine the spatial activation source in both ventricles fully automatically from a standard clinical BSPM and output the result either on a generic mean shape or -- if available -- on the patient-specific geometry. Thus, a patient-specific geometry is not intrinsically needed opposed to many other methods as mentioned above. Our approach can additionally predict if the excitation origin is located on the epicardial or endocardial surface (binary transmural coordinate). We explain how we generated the simulated signals utilized for the training of the CNNs, motivate pre-processing steps of the data, explain how we tackled the problem of inter-patient variability, present the used CNNs and finally demonstrate clinical performance of the method by applying it to real clinical patient signals.

\section{Methods}
\label{meth}

This study consists of two main steps: first, we generated a data set containing more than one million simulated BSPMs by setting different excitation origins in the ventricles. Second, these BSPMs, together with the known origins, were used to train a machine learning pipeline consisting of two CNNs to localize the excitation origin. The trained CNNs were finally applied to unseen clinical data to evaluate their performance with respect to a possible clinical application. In the following, the single steps are described in more detail. After the experiments described in Sections~\ref{meth} and \ref{res}, we added an additional singular value decomposition (SVD) step, which is described in Section~\ref{improv:clinicalData}.

\subsection{Generation of simulated data}
\label{meth:syntheticData}
\begin{figure*}[tb]
\centering
\includegraphics[width=\linewidth]{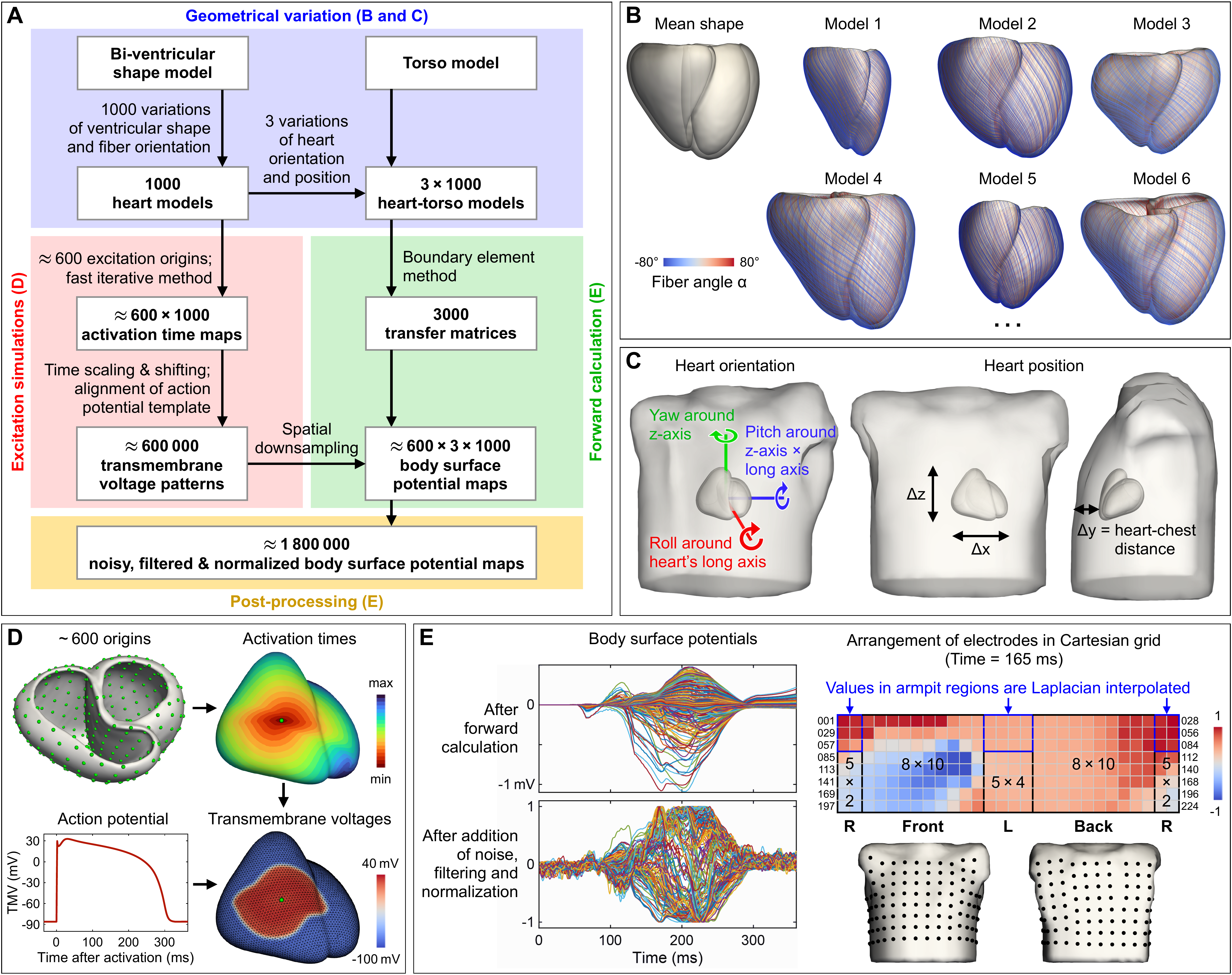}
\vspace{-4mm}
\caption{Workflow to generate simulated data (\textbf{A}) and illustration of the main steps: Variation of ventricular shape and fiber orientation (\textbf{B}), variation of heart orientation and position within the torso (\textbf{C}), excitation propagation simulations (\textbf{D}), forward calculation and processing of body surface potentials (\textbf{E}).}
\label{fig:dataGeneration}
\end{figure*}

The workflow used to generate simulated data is shown in Fig.~\ref{fig:dataGeneration}A. It can be divided into four steps as detailed in the following. In total, 1000 heart models with about 600 excitation origins each were placed at three positions (varying translation and rotation) within a torso model, resulting in a total of about 1.8 million simulated BSPMs.

\subsubsection{Geometrical variation}
\label{meth:geoVar}
As basis for the heart models, we adapted the bi-ventricular statistical shape model (SSM) from~\cite{bai15,demarvao14}, which was created from MR images of over 1000 subjects. It consists of 100 principal components and variances describing the joint variation of disconnected surfaces of the left ventricular (LV) myocardium and the right ventricular (RV) blood pool. To use the SSM for computer simulations, we had to obtain one joint surface of the bi-ventricular myocardium. To this end, we added an orifice at the base of the RV blood pool and shifted the resulting RV endocardial surface by a fixed wall thickness of $3\,\mathrm{mm}$ along its normals to get an RV epicardial surface. The RV surfaces were then merged with the LV surface and the joint surface was remeshed using Instant Meshes~\cite{jakob15} and tetrahedralized using Gmsh~\cite{geuzaine09}. Finally, the principal components were interpolated with Laplacian interpolation~\cite{oostendorp89,jacobson18} to the volume mesh (mean edge length: $0.8\,\mathrm{mm}$). The adapted SSM with ready-to-use instances is available at~\cite{schuler21shapemodel}.

To create the 1000 different heart models, the 100 principal components of the SSM were scaled by uniformly distributed weights in the range $[-3,3]$ standard deviations. A uniform distribution was chosen because we aimed for a good performance also in rare cases, which would be underrepresented using a normal distribution.
Uniform distributions were used for all parameters which were varied during data generation and the scrambled Halton sequence~\cite{kocis97} was used for each group of parameters to get reasonable approximations of multidimensional uniform distributions even with moderate numbers of samples.
For each heart model, different ventricular fiber orientations were assigned using an adapted version\footnote{\href{https://github.com/KIT-IBT/LDRB\_Fibers}{https://github.com/KIT-IBT/LDRB\_Fibers}} of the algorithm proposed in~\cite{bayer12}. The fiber angles $\alpha$ on the epi- and endocardium were varied in the ranges $[-80^\circ,-40^\circ]$ and $[40^\circ,80^\circ]$, respectively~\cite{greenbaum81}. The mean shape and six exemplary heart models including fiber orientations are depicted in Fig.~\ref{fig:dataGeneration}B.

Next, we combined the heart models with a torso geometry to obtain heart-torso models (Fig.~\ref{fig:dataGeneration}C).
We decided to use a single torso geometry because the shape and size of the torso mainly alters the amplitude of the body surface potentials (BSPs) and has only a minor effect on their morphology. It was also shown before that the influence on the simulation results is minor~\cite{Luongo-2020-ID14165}. However, the position and orientation of the heart within the torso has a larger influence~\cite{hoekema99,minchole19} and was therefore varied. The baseline position was determined by aligning the centroid of the mean shape with the centroid of the torso-specific heart. The baseline orientation was determined by the mean angles reported in~\cite[Tab. 1]{odille17}. Three variations of position and orientation pairs were then obtained for each of the 1000 heart models by applying the following geometrical transformations (Fig.~\ref{fig:dataGeneration}C):
\begin{enumerate}
\item Rotation around heart's long axis (roll)
\item Rotation around $z$-axis (yaw)
\item Rotation around cross-product of $z$-axis and heart's long axis (pitch)
\item Translation in $x$-direction ($\Delta x$)
\item Translation in $z$-direction ($\Delta z$)
\item Translation in $y$-direction (heart-chest distance $\Delta y$).
\end{enumerate}
The rotational axes were aligned with the heart's centroid, i.e. all axes intersect in the centroid.
The 6 parameters were varied across all 3000 heart-torso models. The ranges were $[-20^\circ,20^\circ]$ for the roll, yaw and pitch angles~\cite{odille17}, $[-20\,\mathrm{mm},20\,\mathrm{mm}]$ for $\Delta x$ and $\Delta z$~\cite{curtin17} and $[12\,\mathrm{mm},52\,\mathrm{mm}]$ for $\Delta y$~\cite{rahko08}.
Intracavitary blood was included in each heart-torso model by extracting and closing the endocardial surface as it is one of the most important torso inhomogeneities regarding the effect on the ECG~\cite{brody56,vanoosterom89}.

\subsubsection{Excitation propagation simulations}
\label{meth:excSim}
Excitation propagation simulations (Fig.~\ref{fig:dataGeneration}D) were performed with an anisotropic eikonal model as used in~\cite{Schuler_2022}, which is computationally less demanding than a monodomain model but can reproduce the main characteristics of cardiac excitation spread~\cite{Franzone1993}. The conduction velocity (CV) anisotropy ratio was set to $2.7$~\cite{pop12} and the baseline CV in fiber direction was set to $0.6\,\mathrm{m/s}$~\cite{kleber11}.
Approximately 600 excitation origins (588.802 on average) were uniformly distributed over the epi- and endocardial surfaces of each heart model~\cite{jakob15}. For each excitation origin, the eikonal equation was solved using a GPU implementation of the fast iterative method\footnote{\href{https://github.com/KIT-IBT/FIM\_Eikonal}{https://github.com/KIT-IBT/FIM\_Eikonal}}~\cite{fu13}. Then, each of the resulting 588\,802 activation time (AT) maps was scaled by a different factor in the range $[0.5,1.5]$, which corresponds to CVs in the range $[0.4\,\mathrm{m/s},1.2\,\mathrm{m/s}]$. Furthermore, AT maps were shifted in the range $[0,100\,\mathrm{ms}]$, because the precise start of activation is unknown in practice. An action potential template based on the model by ten Tusscher et al.~\cite{tentusscher06} was then aligned with the ATs to obtain transmembrane voltages (TMV) at a sampling rate of $1\,\mathrm{kHz}$.

\subsubsection{Forward calculation}
\label{meth:fwdCalc}
Fast forward calculations were achieved by computing transfer matrices for each heart-torso model using the boundary element method~\cite{stenroos07,simms95}. To this end, the original heart surfaces were remeshed with a reduced number of $12\,\mathrm{k}$ nodes (mean edge length: $2.4\,\mathrm{mm}$) and the TMVs were  downsampled with Laplacian downsampling~\cite{schuler19} to the reduced mesh. Tissue conductivities were assumed as isotropic and set to $0.05\,\mathrm{S/m}$ and $0.15\,\mathrm{S/m}$ in the intra- and extracellular domains of the heart, $0.6\,\mathrm{S/m}$ in the blood and $0.2\,\mathrm{S/m}$ in the rest of the torso.
BSPs were computed at 200 electrodes (black dots in Fig.~\ref{fig:dataGeneration}E).

\subsubsection{Processing simulated BSPMs}
\label{meth:postProc}
To represent clinically occurring artifacts in the BSPs in order to minmize the domain gap between in silico and in vivo signals, we added $30\,\mathrm{dB}$ of white Gaussian noise, $30\,\mathrm{dB}$ of recorded muscle artifacts and $10\,\mathrm{dB}$ of recorded baseline drift (signal-to-noise ratios refer to the average over all electrodes during the activation). The latter two noise signals were obtained from~\cite{moody84}. Then, the noisy BSPs were filtered with clinical settings (fourth order Butterworth bandpass, $0.5\text{--}150\,\mathrm{Hz}$). To reduce the influence of torso size and absolute conductivities, the BSPs were normalized by dividing by their maximum absolute value over time for each channel separately.
Finally, the BSPs were arranged in a Cartesian grid and missing values at the armpits were interpolated (Laplacian interpolation with periodic boundary conditions in horizontal direction)~\cite{knyazev15}. This approach results in a total of $8\times28=224$ BSP values per time step.
The full BSPs (including repolarization) were truncated at the end because the activation is sufficient for localization of the excitation origin. As the precise end of activation is unknown in practice, the truncation was applied at varying times in the range $[0,100\,\mathrm{ms}]$ after the true end. 
Fig.~\ref{fig:dataGeneration}E shows the BSPs corresponding to the excitation in Fig.~\ref{fig:dataGeneration}D after forward calculation, after normalization and after arrangement in the Cartesian grid.

\subsection{Machine learning pipeline}
\label{meth:mlPipeline}
\begin{figure*}[tb]
\centering
\includegraphics[width=\linewidth]{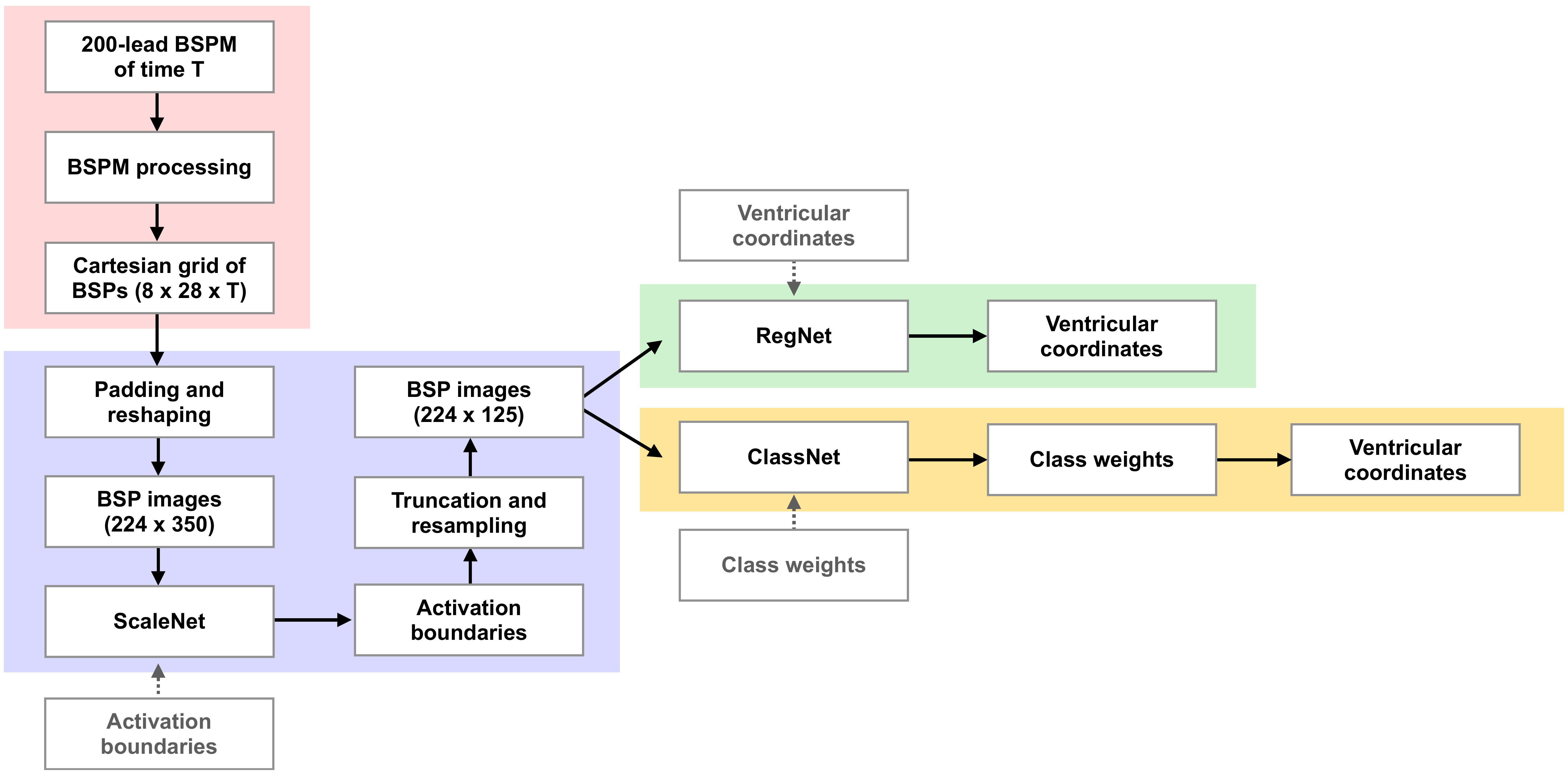}
\vspace{-4mm}
\caption{The two-step approach for the localization of the excitation origin. Red: simulated BSPMs were processed and arranged into a Cartesian grid with their individual length T (Sec.~\ref{meth:postProc}). Blue: Afterwards, the signals were padded and reshaped to generate BSP images that were used for ScaleNet. The predicted activation boundaries were used to truncate and resample the BSP images. Green: The cropped and resampled BSP images were analyzed with RegNet using the known ventricular coordinates for the training. Yellow: ClassNet used the same BSP image data as RegNet, however, RegNet directly outputs ventricular coordinates whereas ClassNet outputs class weights (and thereby indirectly a likelihood). The class weights are then converted to coordinates. The light grey boxes contain the respective input labels necessary for the training of the respective networks.}
\label{fig:mlpipeline}
\end{figure*}

A two-step approach was implemented to localize the excitation origins from the BSPMs (Fig.~\ref{fig:mlpipeline}): First, a CNN (ScaleNet, in the blue part of Fig.~\ref{fig:mlpipeline}) was trained to detect the start and end of the relevant activation window (activation boundaries) in the BSPMs. We decided to train this dedicated CNN explicitly separate from the following localization CNN for the following reasons: We assumed that the depolarization phase is the most relevant part in the ECG for the detection of the excitation origin. The timing of the depolarization of the ventricles, however, differs slightly from the QRS complex that is visible in the ECG/BSPM signal as very early and late activations (mostly with a low spatial extent) in the heart might be dominated by noise in the surface signals. With the known activation start and end times from the simulation, which are the earliest and latest activation times in the heart, the CNN was able to learn to detect the real start and end times from the surface signals. By introducing this separate network, we were able to easily incorporate our assumption of the relevant parts for the localization into the learning approach. This CNN could then be applied to clinical data, too. A further reason for the introduction of ScaleNet is related to the learning speed of CNNs. With BSPMs, cropped  to the start and end of the activation and thus, containing only relevant information, the training of the subsequent methods was accelerated. Furthermore, we were able to separate the effects of the single tasks (activation detection and localization) on the final results.

With a cropped version of the initial BSPMs and the known excitation origins from the simulations, two different localization CNNs were trained. We implemented and compared both a regression-based approach (RegNet, green part of Fig.~\ref{fig:mlpipeline}) directly delivering the coordinates of the excitation origin and a fuzzy classification approach (ClassNet, yellow part of Fig.~\ref{fig:mlpipeline}) delivering class weights that could then be converted to ventricular coordinates in an additional step.

For the training of the methods, the data set with the simulated BSPMs was split into a training, validation and testing set according to the 1000 unique heart models. The training set contained 700 heart models, validation and testing set 150 each. Thus, each set contained new, unseen ``virtual patients''. The available clinical data set was used only for testing and not considered during training and validation.

All three CNNs (ScaleNet, RegNet and ClassNet) are based on the ResNeXt-50 architecture~\citep{Xie2016}. The CNNs were intialized according to He~\citep{He2015Init} and trained using the Adam optimizer~\citep{Kingma2014}. The learning rate was cyclically adjusted according to the approach proposed by Smith et al.~\citep{Smith2017}. The minimum and maximum learning rate were determined with a learning rate range test adapted from~\cite{Smith2017}. However, we determined the learning rate range not by visual analysis as originally proposed but by fitting a polynomial function and using the concept of the L-curve~\citep{Hansen1993}. Dropout with a rate of 0.2 was implemented in the last two (fully connected) layers of the CNNs. Early stopping was used for all trainings: the validation loss was monitored during the training and the stopping epoch was chosen when the validation loss increased over 5 epochs. All CNNs were realized using PyTorch~\citep{Paszke2019}.

As described, the underlying network structures and training procedures were similar for all three CNNs. Nevertheless, there were CNN-specific parameters, which are described in the following sections.

\subsubsection{Time scaling (ScaleNet)}
\label{meth:timeScaling}
As usual for regression tasks, the mean squared error loss was used as loss function for the training of the ScaleNet CNN. The first layer of the standard ResNeXt-50 architecture was changed to accept images of size 224~$\times$~350 (leads~$\times$~samples). To be able to use the simulated data with this input layer, the BSPMs (which were of different size initially) were padded to a length of 700 time steps by repeating the last time step and were then downsampled from 1000\,Hz to 500\,Hz yielding the desired image size. The last layer of the network architecture was adapted to deliver two outputs: the start and the end time of the ventricular activation. 

\subsubsection{Regression (RegNet)}
\label{meth:regression}
\begin{figure}[htb]
\centering
\includegraphics[width=\linewidth]{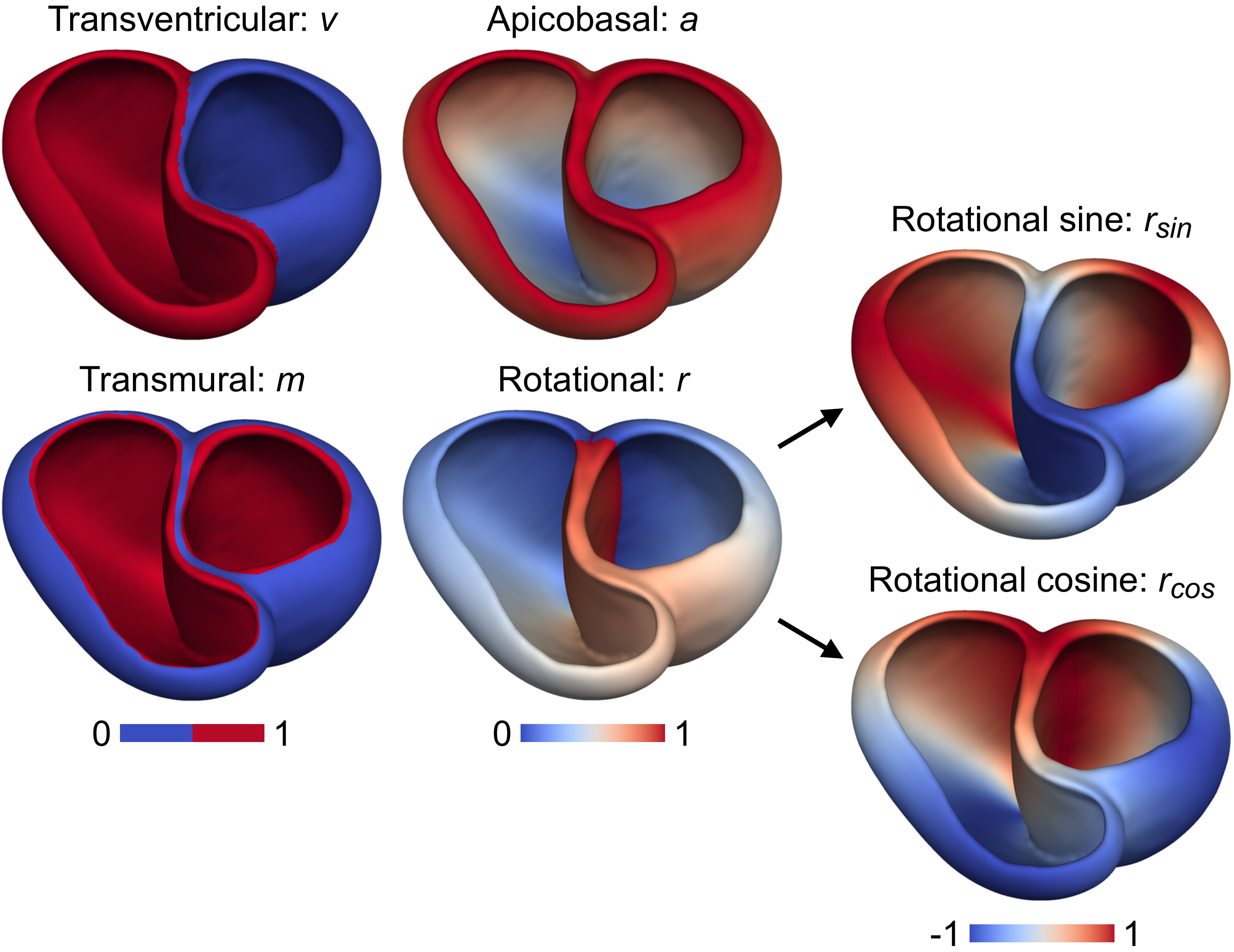}
\caption{Ventricular coordinates according to Cobiveco, used for regression and mapping. The rotational coordinate was replaced by its sine and cosine to prevent discontinuous jumps in the coordinate space.}
\label{fig:cobiveco}
\end{figure}

The localization of the excitation origin and its visualization highly depends on the local coordinate system of the patient geometry. Without a prior registration, it is hardly possible to train a CNN with Euclidean coordinates. To overcome this, we converted the Euclidean coordinates describing the excitation origins used in the simulations into a parameterized form using the Cobiveco approach~\citep{schuler21cobiveco}. Cobiveco describes a ventricular position by four normalized coordinates as shown in Fig.\ \ref{fig:cobiveco}: binary transventricular and transmural coordinates $v$ and $m$ and continuous apicobasal and rotational coordinates $a$ and $r$. The latter, however, shows a jump from 1 to 0 at the posterior junction between the septum and the free walls. To prevent problems during training, we replaced the rotational coordinate by its sine and cosine. This yields a continuous description of the rotational position without loosing the uniqueness of the coordinate. Regarding the transmural coordinate $m$, which is continuous in the original Cobiveco formulation, it should be noted that in our simulated data set, only discrete values of 0 and 1 exist, as clinically there is usually only the distinction between endo- and epicardial. By using these parameterized coordinates, we are able to display positions obtained from our methods not only on the patient-specific geometry but also on a mean shape (or every other geometry). By doing so, results can always be visualized at least on the mean shape even if a patient-geometry is not available. This might help physicians to better locate possible ablation sites even without a prior imaging step.

The input layer of the standard ResNeXt-50 architecture was adjusted to images of size 224~$\times$~125 (leads~$\times$~samples). This corresponds to the size of the images after the cropping with the results from ScaleNet and a resampling to 125 samples.

In total, RegNet had to output five coordinates (Fig.~\ref{fig:cobiveco}), some of them continuous and some binary. This is why we implemented a multi-task loss consisting of three mean squared errors (MSEs) for the continuous coordinates and two binary cross entropies (BCEs) for the binary coordinates. All five error terms were weighted based on coordinate velocities and empirical findings, e.g. the rotational error was weighted with the square root of the apicobasal coordinate, since the rotational coordinate velocity is dependent on the apicobasal position. This led to the loss function:
\begin{align}
\label{eqn:LossRegressionNet}
L_\text{RegNet} &=\frac{1}{N}  \sum^N_{n=1}\bigl( \text{MSE}_{a,n} + \text{MSE}_{r\_sin,n} + \text{MSE}_{r\_cos,n} \nonumber\\
&\phantom{=}\ + \text{BCE}_{m,n} + \text{BCE}_{v,n}\bigr)
\end{align}
with
\begin{align}
\text{MSE}_{a,n} &= \bigl(2.5\,(l_{a,n} - y_{a,n})\bigr)^2 \ ,\\
\text{MSE}_{r\_sin,n} &= \bigl(\sqrt{l_{a,n}}\,(l_{r\_sin,n} - y_{r\_sin,n})\bigr)^2 \ ,\\
\text{MSE}_{r\_cos,n} &= \bigl(\sqrt{l_{a,n}}\,(l_{r\_cos,n} - y_{r\_cos,n})\bigr)^2 \ ,\\
\text{BCE}_{m,n} &= l_{m,n}\,\ln\bigl(\sigma(y_{m,n})\bigr) \nonumber \\&\phantom{=}\ +(1-l_{m,n})\,\ln\bigl(1-\sigma(y_{m,n})\bigr) \ ,\\
\text{BCE}_{v,n} &= l_{v,n}\,\ln\bigl(\sigma(y_{v,n})\bigr) \nonumber \\&\phantom{=}\ +(1-l_{v,n})\,\ln\bigl(1-\sigma(y_{v,n})\bigr) \ ,
\end{align}
where $l$ denotes the labels, $y$ the predictions and $N$ the number of samples that are evaluated. The indices refer to the different Cobiveco componentes. The sigmoid function $\sigma$ keeps the argument of the logarithm above zero.

The last layer of the standard ResNeXt-50 architecture was accordingly changed to have five outputs.

\subsubsection{Fuzzy classification (ClassNet)}
\label{meth:classification}
\begin{figure}[htb]
\centering
\includegraphics[width=\linewidth]{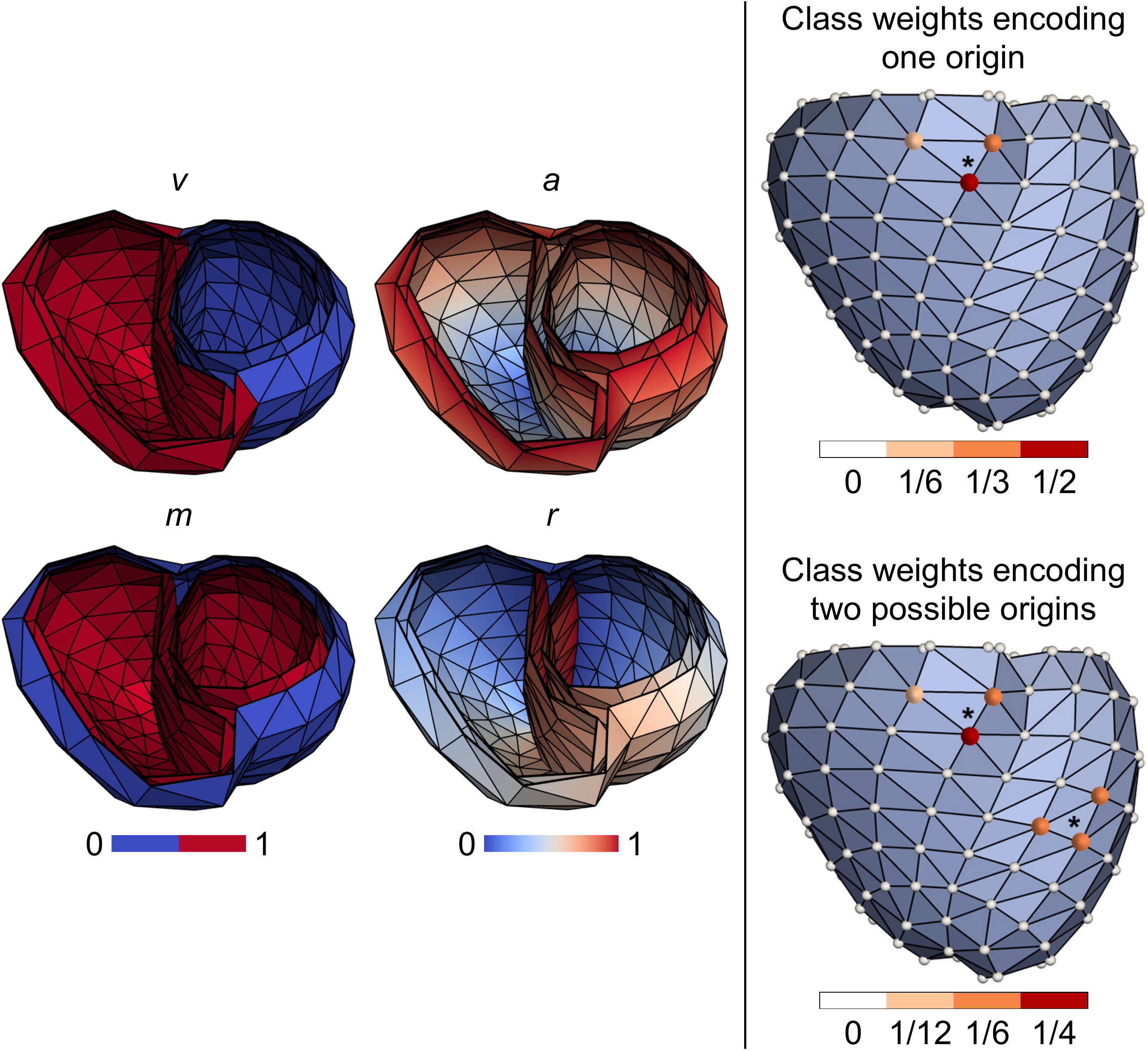}
\caption{Coarse mesh with ventricular coordinates (\textit{left}) used to obtain 303 barycentric coordinates (or ``class weights'') (\textit{right}) for classification. The asterisks denote the estimated origins based on the class weights.}
\label{fig:classWeights}
\end{figure}

The regression approach as presented in Sec.~\ref{meth:regression} yields exactly one solution. However, dealing with an inverse problem, we have to assume that multiple origins can deliver similar surface signals. Thus, if the regression approach does not predict the correct position, there is no hint for the physician where to look next during an ablation procedure. Since there is a lot of recent work about uncertainty quantification for classification~\cite{Abdar_2021}, we tried to tackle this problem by converting the regression problem into a classification problem. A common approach to do this is to discretize the ventricles into segments and then let a classification technique predict the segment containing the excitation origin instead of estimating the exact position. However, by doing so the spatial resolution is markedly reduced. Therefore, we introduced a fuzzy classification scheme to be able to translate the class weights back to an exact position on the ventricular surface. We coarsened a mean ventricular surface triangular mesh to 540 triangles with 303 vertices as shown in Fig.~\ref{fig:classWeights}. In this coarse mesh, every triangle either belonged to the LV or RV, as well as either to the endo- or epicardium (Fig.~\ref{fig:classWeights} left). The vertices formed the classes of the classification problem. The class values of the excitation origins used during training were obtained by calculating the barycentric coordinates of the excitation origin in the surrounding triangle (for an example, see Fig.~\ref{fig:classWeights} upper right). By definition, the sum of the coordinates is normalized to one and the single (vertex) coordinate values were stored as three of the 303 class values. All other 300 elements of remained 0 for the respective experiment (combination of one heart-torso model and one excitation origin). Thus, our labels are not one-hot encoding, opposed to most common classification tasks. As a reminder, every class has a corresponding vertex containing one coordinate value. And to be clear again, every experiment is transferred to the same (mean) geometry by using the Cobiveco coordinates to ensure consistent classes and to have one general mesh.

To be able to later return from the class values to the location of the excitation origin, we converted the class values that were the output of the classification network back to a position on the surface defined by the ventricular coordinates. However, it is expected that the network predicts the values of more than three classes to be larger than zero. In this case, the sum of the class values (vertices) connected to one triangle can be interpreted as a kind of ``probability'' or better likelihood\footnote{We want to emphasize that the outputs of the CNN do not correspond to a realistic, trustworthy probability in a statistical sense~\cite{Gal2016}. CNNs are known to be over-confident~\cite{Guo2017}, hence, the absolute value of the likelihood does not reflect the actual probability for a solution. This is why we stick with the term ``likelihood'' which we understand in a sense of solution with highest plausibility based on knowledge from the data.} and thus we can sort the single sums of class values by their value to obtain potential solutions ordered by likelihood. An example for two different solutions is shown in Fig.~\ref{fig:classWeights} (lower right corner). Here, there are two triangles, each formed by three class weights (vertices) having a non-zero value. These could be interpreted as two different possible excitation origins. With this problem formulation, we could overcome the drawback of decreased spatial resolution by the introduction of ventricular classes. Additionally, it is a first step towards predicting several ranked possible solutions, giving a physician multiple possible solutions to have a closer look at.

Having these 303 barycentric coordinates (or ``class weights'', one per vertex) available for ClassNet, we needed to change the last layer of the ResNeXt-50 architecture to have 303 output neurons. The input layer was the same as for RegNet (Sec.~\ref{meth:regression}). As loss function, we chose the categorical cross entropy (cross entropy of softmax of network output):
\begin{equation}
\label{eqn:LossFuzzyNet303}
L_{\text{ClassNet}}=\frac{1}{N} \sum_{n=1}^{N} \left( -\sum_{i=1}^{303} l_{i,n} \ln\left(  \frac{e^{y_{i,n}}}{\sum_{j=1}^{303} e^{y_{j,n}}} \right)\right) \text{  ,}
\end{equation}
again with the labels $l$, the predictions $y$, the number of samples to be evaluated $N$, but now containing class weights.

\subsubsection{Evaluation on simulated data}
\label{meth:evalSynthetic}
The simulated test set contained 264\,978 cases from 150 heart models that were not included during training or validation. The accuracy of the ScaleNet predictions was evaluated in terms of absolute errors with respect to the true start and end of activation. For RegNet and ClassNet, we used the model-specific cardiac geometries to compute the geodesic distance to the true origin through the myocardial volume. Furthermore, we report the transventricular and transmural misclassifications.

\subsubsection{Evaluation on clinical data}
\label{meth:evalClinical}
The clinical test data comprised 67 cases from 37 cardiac resynchronization therapy (CRT) patients.
This data set was acquired in clinical studies~\citep{revishvili15a,chmelevsky2018} at the Almazov National Medical Research Center in Saint Petersburg (Russia), which adhered to the Declaration of Helsinki and were approved by the local institutional review board. Written informed consent was obtained from each patient.
BSPs were recorded during isolated LV or RV pacing from the electrodes of an implanted pacemaker.
Although only the positions of ECG electrodes are required to apply our method, computed tomography (CT) images of the torso and heart were obtained from each patient and were segmented with the software of the Amycard 01C EP system (EP Solutions SA, Yverdon-les-Bains, Switzerland).

\begin{figure}[htb]
\centering
\includegraphics[width=\linewidth]{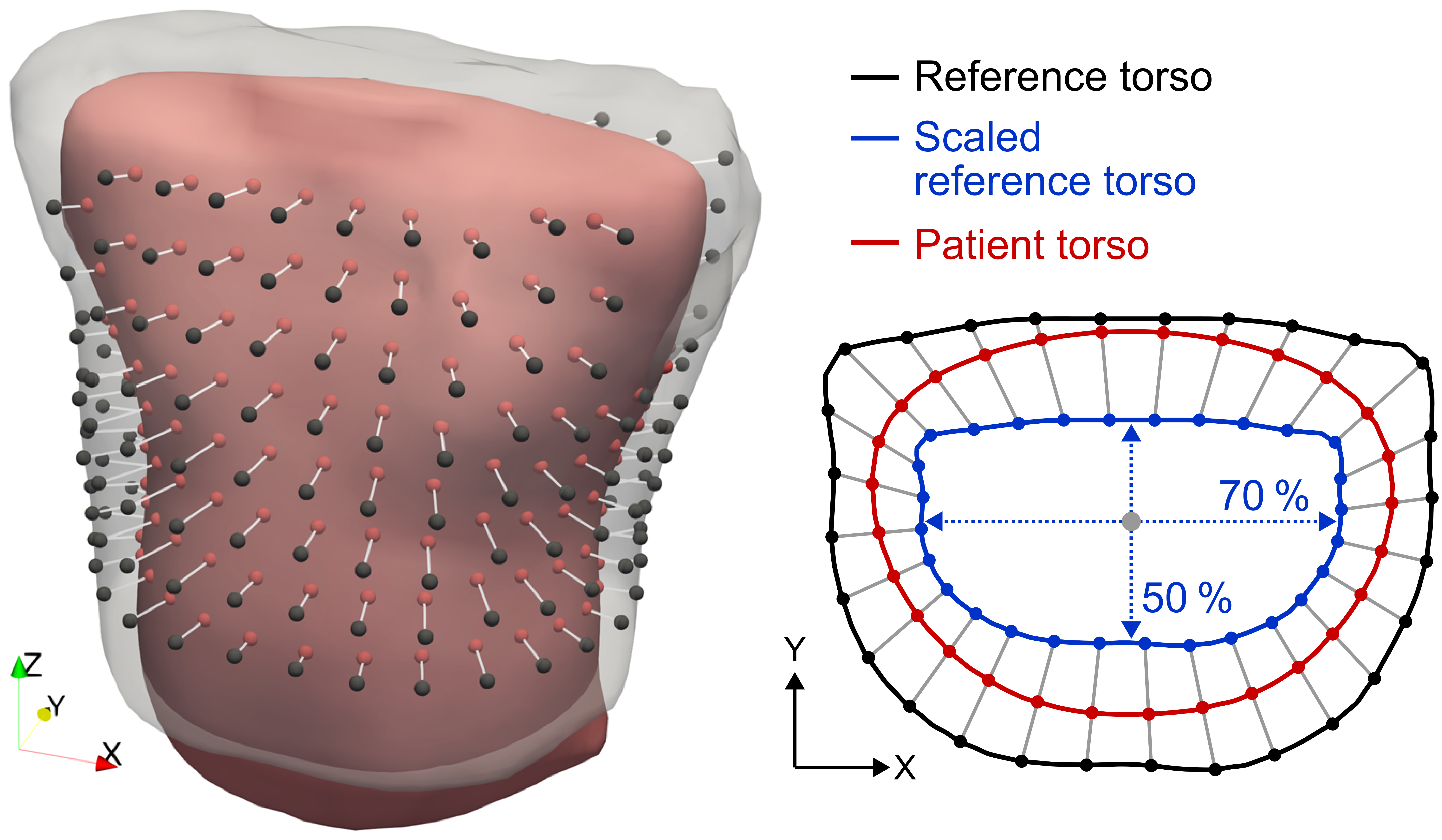}
\caption{Projection of electrodes from the reference torso (black) onto the patient torso (red). The directions of projection (white lines on the \textit{left}; gray lines on the \textit{right}) were determined by scaling the reference torso in $x$- and $y$-directions (blue) and connecting the corresponding electrode positions before and after scaling.}
\label{fig:elecProj}
\end{figure}

In this study, the ECG electrodes were localized from the torso CT. To obtain BSPs at the electrode positions used for the ML pipeline, the electrodes on the reference torso were projected onto the patient torso and the measured BSPs were interpolated with Laplacian interpolation to the projected electrode positions. The projection of electrodes is illustrated in Fig.~\ref{fig:elecProj}. First, the torso surfaces were aligned in $x$- and $y$-directions based on their centroids and in $z$-direction based on the vertical position of the heart. Then, the directions for projection were obtained by scaling the reference torso with respect to its centroid (by $70\,\%$ in $x$- and $50\,\%$ in $y$-direction, reflecting the width/thickness ratio of the torso). This procedure is illustrated in the right panel of Fig.~\ref{fig:elecProj} and was found to be more robust than using local surface normals.

The cardiac CT was used to determine the surface of the ventricular myocardium and the position of the pacemaker electrodes. The latter served as true excitation origin. As for the evaluation on the simulated test data, the geodesic distance to the true origin was computed on the patient-specific cardiac geometry. Unfortunately, no evaluation of the ScaleNet predictions was possible for the clinical data set because the true start and end of activation are unknown. We want to emphasize again that we did not use any clinical data during training but only for validation of the localization CNNs.

\section{Results}
\label{res}
\begin{table}[]
\caption{Localization errors in mm for the simulated test data set and for the clinical data set. Additionally, results for the experiment with an additional singular value decomposition (SVD) step are given. Results are given as median$\pm$interquartile range. The best-performing method is highlighted in bold.}
\label{tab:results}
\begin{tabular}{l|cccc}
                    & ClassNet  & RegNet      & ClassNet & RegNet \\
       &              &          & +SVD      & +SVD \\       
                                 \hline
Test  & \textbf{1.5}$\pm$\textbf{1.3} &  2.3$\pm$1.9   & 2.0$\pm$1.7      & 2.6$\pm$2.2 \\
Clinical    &47.0$\pm$52.2  &  37.0$\pm$30.8   & 40.3$\pm$47.1 & \textbf{32.6}$\pm$\textbf{25.3}
\end{tabular}
\end{table}

\begin{table}[]
\caption{Correct rates for the classification of the binary ventricular coordinates transmural (m) and transventricular (v) for the simulated test data set and for the clinical data set. Additionally, results for the experiment with an additional singular value decomposition (SVD) step are given. The best-performing method is highlighted in bold if there is one.}
\begin{tabular}{l|cccc}
                & ClassNet & RegNet & ClassNet & RegNet \\
                &  & & +SVD & +SVD \\
                \hline
m test     & >0.99    & >0.99       & >0.99  & >0.99      \\
v test     & >0.99    & >0.99        & >0.99  & >0.99      \\
m clinical & 0.64     & 0.67         & 0.63   & \textbf{0.82}       \\
v clinical & 0.67     & 0.72 & 0.82   &  \textbf{0.85}
\end{tabular}
\end{table}

\begin{figure}[tb]
\centering
\includegraphics[width=\linewidth]{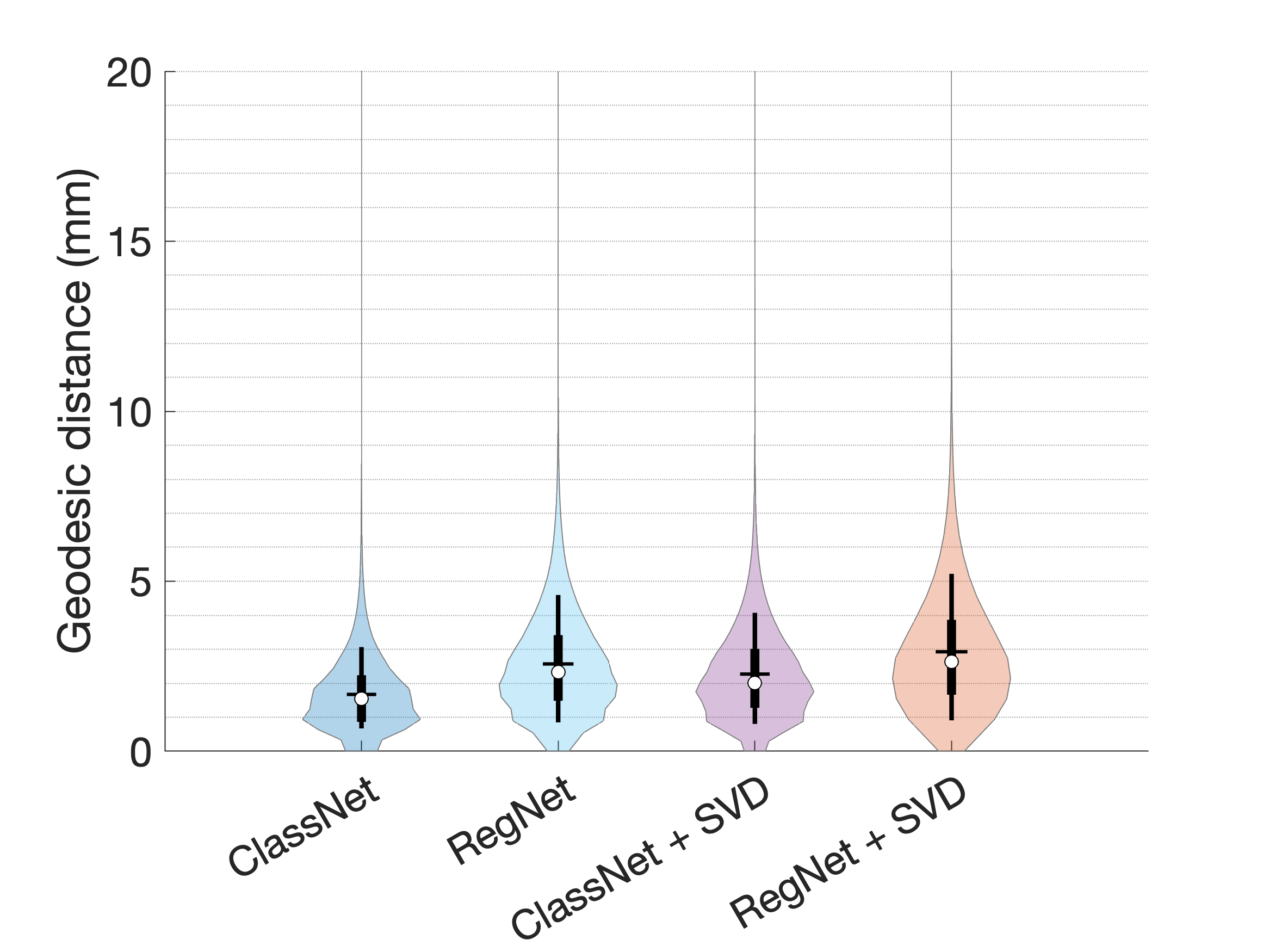}
\caption{Geodesic errors in mm of the localization CNNs on the simulated test data. Violin plots are capped at 20\,mm for better readability and visualize the distribution of the errors with the parameters median (white dot), interquartile range (bold black vertical line) and the 10\textsuperscript{th} and 90\textsuperscript{th} percentile interval (narrow black vertical line) of RegNet and ClassNet, both without and with the SVD step. The horizontal black line represents the mean of the errors.}
\label{fig:geo_test}
\end{figure}

\begin{figure}[tb]
\centering
\includegraphics[width=\linewidth]{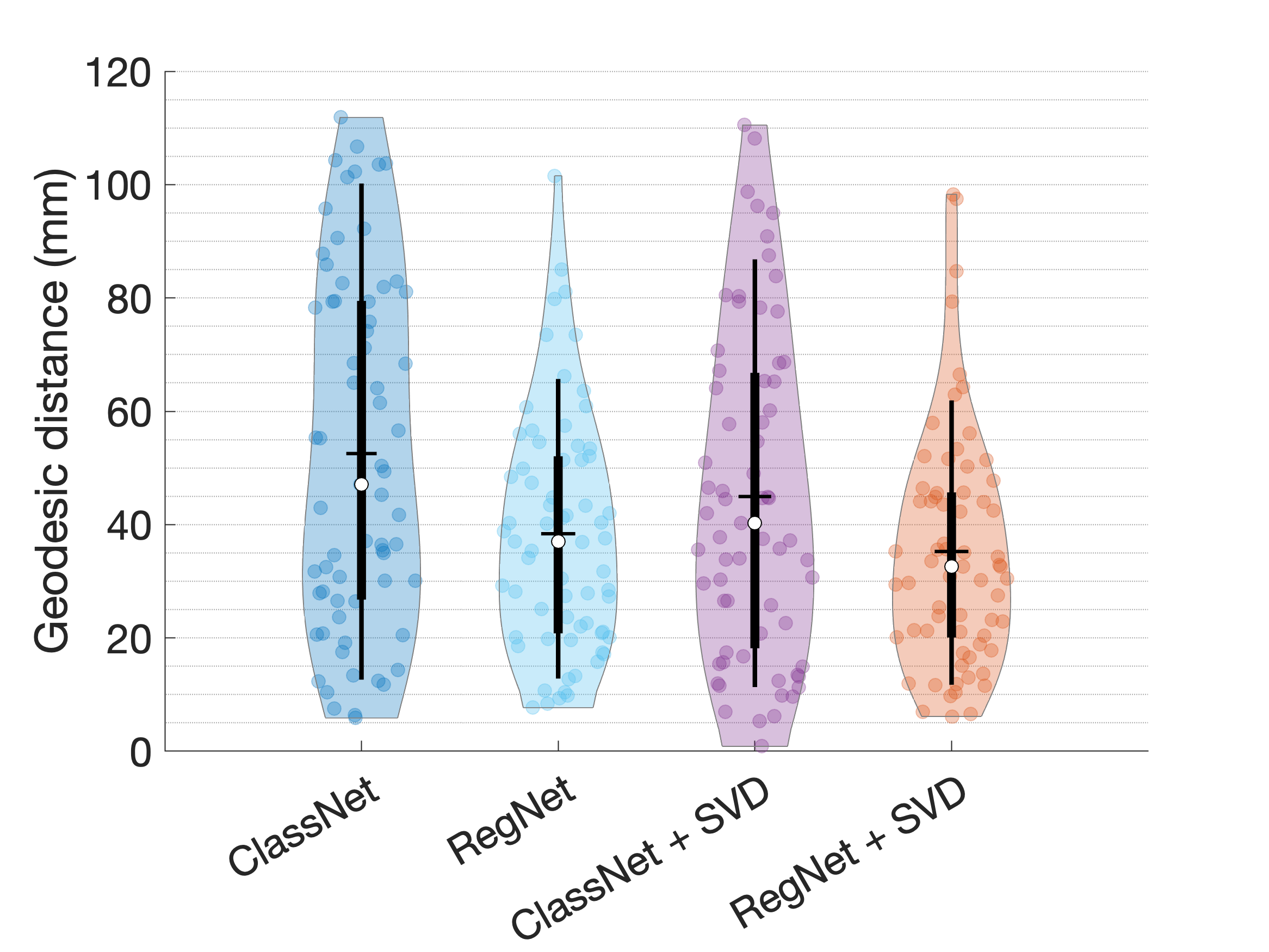}
\caption{Geodesic errors in mm of the localization CNNs on the clinical data. The violin plots visualize the distribution of the errors with the parameters median (white dot), interquartile range (bold black vertical line) and the 10\textsuperscript{th} and 90\textsuperscript{th} percentile interval (narrow black vertical line) of RegNet and ClassNet, both without and with the SVD step. The horizontal black line represents the mean of the errors. The single samples (error values) are shown by the colored circles.}
\label{fig:geo_clinical}
\end{figure}

\subsection{ScaleNet}
\label{res:ScaleNet}
ScaleNet was trained for 40 epochs. The early stopping criterion was met after 32 epochs, so the network was evaluated with the state at epoch 32. Results are given as median$\pm$interquartile range and reported for the prediction of the start and end times separately: for the start, the absolute error was 0.229\,ms$\pm$0.293\,ms and for the end 0.302\,ms$\pm$0.422\,ms on the simulated test data. Since there was no ground truth, ScaleNet could not be validated on the clinical data set.

\subsection{RegNet and ClassNet}
\label{res:localizationNets}
\begin{figure}[tb]
\centering
\includegraphics[width=\linewidth]{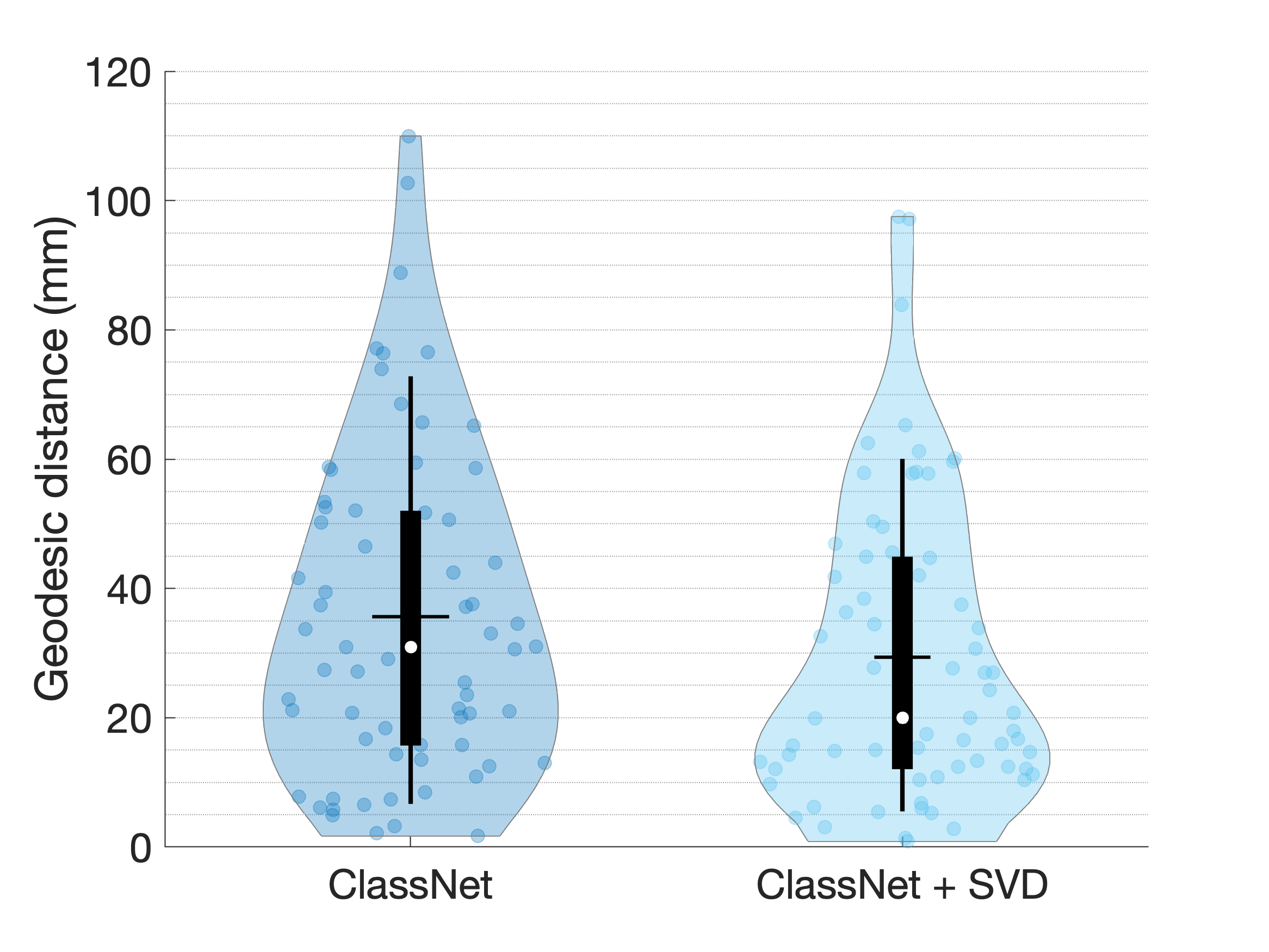}
\caption{The minimum of the geodesic errors in mm of ClassNet's top-3 most likely localization predictions on the clinical data set. The violin plots visualize the distribution of the errors with the parameter median (white dot), interquartile range (bold black vertical line) and the 10\textsuperscript{th} and 90\textsuperscript{th} percentile interval (narrow black vertical line), without and with the SVD step. The horizontal black line represents the mean of the errors. The single samples (error values) are shown by the colored circles.}
\label{fig:geo_top3_clinical}
\end{figure}

\begin{figure}[tb]
\centering
\includegraphics[width=0.7\linewidth]{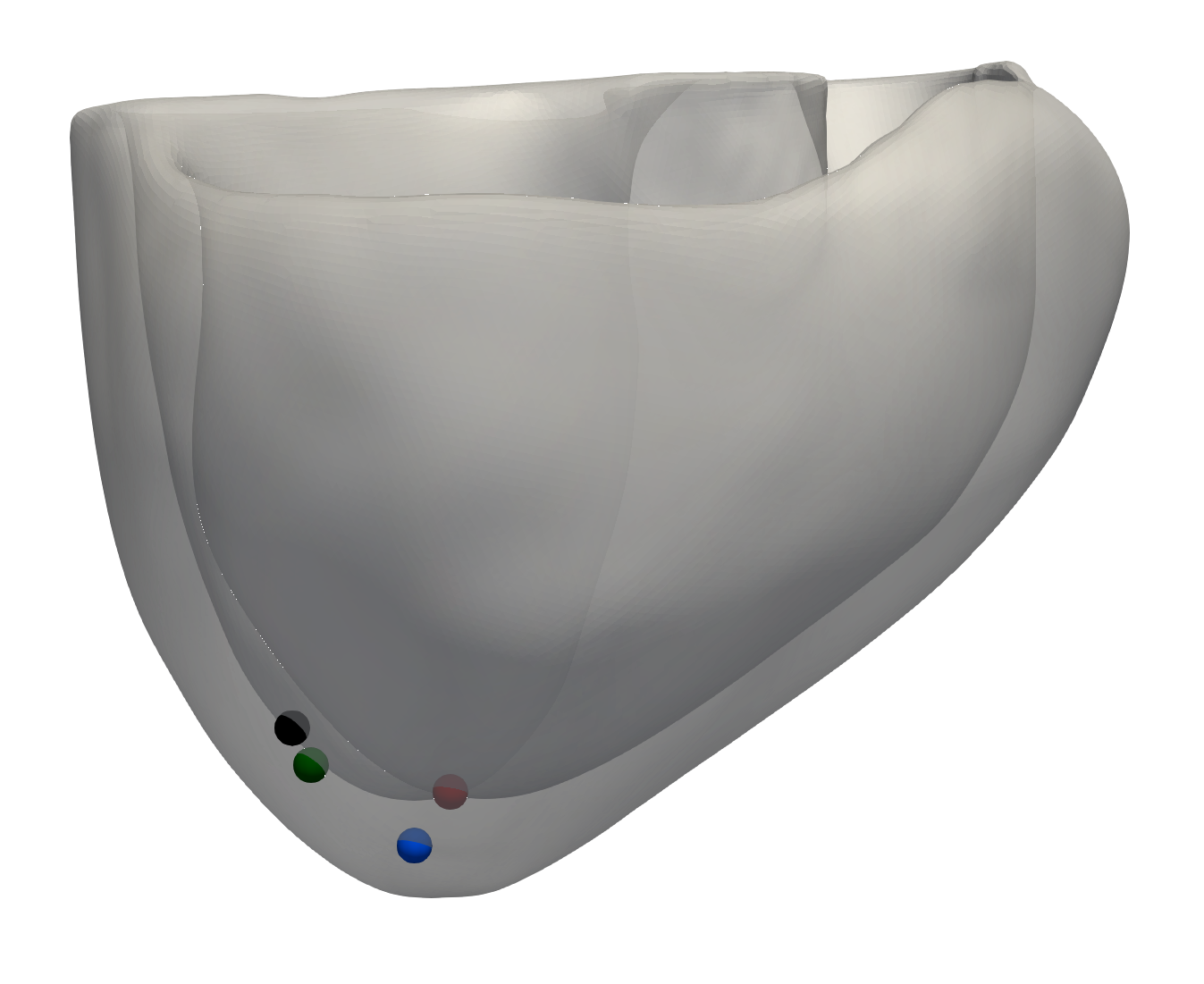}
\caption{One clinical geometry, the three most probable localization results from ClassNet (without SVD) for this patient and the ground truth. The black dot visualizes the true excitation origin, the red dot represents the most likely solution (in this case, it is the only one located on the endocardium). The green and blue dots represent the second and third most likely, respectively. In this case, the second guess reduced the localization error from 23.0\,mm to 7.7\,mm.}
\label{fig:example_clinical}
\end{figure}

RegNet and ClassNet were trained for 47 and 41 epochs, respectively, with the early stopping criterion being met after 39 and 15 epochs. Again, the network weights for evaluation were chosen from these epochs. Localization errors were evaluated by calculating the geodesic instead of the often used Euclidean distances between estimated and true excitation origin. The localization errors of RegNet and ClassNet are given in the first two columns of Tab.~\ref{tab:results} for the simulated test data and the clinical data, as well as in the form of the first two violin plots in Figs.~\ref{fig:geo_test} and \ref{fig:geo_clinical}. Both localization CNNs detected the excitation origins with a median error of below 2.5\,mm. However, ClassNet performed better than RegNet regarding median and interquartile range. The overall distribution of errors for ClassNet tended to lower values than the distribution of RegNet (Fig.~\ref{fig:geo_test}). In contrast, RegNet outperformed ClassNet on the clinical data set. Both CNNs did not see the clinical data set during training. Here, median errors ranged up to 37\,mm-47\,mm. Both the mean and maximum localization errors of RegNet were smaller than using ClassNet (Fig.~\ref{fig:geo_clinical}). The previously discussed possibility to interpret the output of the network as multiple possible solutions enables us to calculate a top-3 error by selecting the three solutions with the highest triangle weights and choosing the smallest localization error among them. For this, the neighboring triangles of already found solutions were excluded since the sum of the vertex values might also be large just by containing the same vertex connected to a high class value\footnote{The problem gets apparent when looking at Fig.~\ref{fig:classWeights} upper right. The vertex with the value (class weight) 1/2 is also part of the surrounding triangles. This leads to a high sum of vertex values for the surrounding triangles but not offering a distinct alternative solution.}. For the clinical data, this leads to the results shown in the left violin plot of Fig.~\ref{fig:geo_top3_clinical}. It can be observed that, compared to Fig.~\ref{fig:geo_clinical}, the median error decreased from 47.0\,mm to 30.9\,mm, whereas the interquartile range dropped from 52.2\,mm to 36.2\,mm (compare Tab.~\ref{tab:results}). One example visualizing the three excitation origins with the highest triangle weights and the true location is shown in Fig.~\ref{fig:example_clinical}.

\section{Improving results on clinical data with SVD filtering}
\label{improv:clinicalData}
By comparing the results on the simulated and the clinical test data sets, a domain gap becomes apparent which the CNNs could not close by generalization. As realistic noise was added to the simulated signals, we considered simplifications in the excitation propagation model and forward model the most likely reason.

These idealized simulations lead to details in the simulated BSP images that are not visible in the clinical data. If the CNNs now ``focus'' on these details during training, the localization procedure might fail on clinical data since these details are not present there. This is why we looked for a way of regularization that prevents the network to ``look'' on these details. Instead of considering this during training, we processed the data in a way such that details that are not present in most of the data are blanked. Such controlled information reduction can be realized by SVD filtering. The SVD was calculated with the training data after resampling and cropping based on ScaleNet's predictions. The basis that was computed in this way was used to filter also the simulated validation data, the simulated test data and the clinical data. RegNet and ClassNet were re-trained with the cropped, scaled and SVD-filtered BSP images using the same training strategy as explained in Sec.~\ref{meth:mlPipeline}. The SVD step improved median localization errors on clinical data by 4 to 7\,mm (Tab.~\ref{tab:results}, Fig.~\ref{fig:geo_clinical}).  While results on the clinical data improved, the distribution of error on the simulated data widened and higher maximum errors occurred (Fig.~\ref{fig:geo_test}). The statistical performance measures on the simulated test data worsened only slightly by a maximum of 0.5\,mm though (Tab.~\ref{tab:results}). The median top-3 error of the three solutions with the highest likelihood obtained with ClassNet on the clinical data set (right violin plot of Fig.~\ref{fig:geo_top3_clinical}) dropped from 32.6\,mm to 20.0\,mm, the interquartile range from 47.1\,mm to 32.6\,mm.

\section{Discussion}
\label{disc}
In this work, we presented recent advances in the field of non-invasive localization of cardiac excitation origins. We presented two methods that are capable of fully-automatically localizing the excitation origin from BSPMs without the need of a patient-specific geometry. One of those is even capable of detecting several possible excitation origins and ranking them. The underlying CNNs were trained only on simulated data, however directly tested on clinical data. To our knowledge, there has never been a method proposed offering the bi-ventricular localization without a patient-specific geometry, being evaluated directly on clinical data after training with simulated data, and by method selection inherently offering several possible localization results. A method like this can save time and money as prior tomographic scans can be skipped and the results based on the CNNs are available in seconds and make the localization accessible to clinical sites without imaging capabilities. Besides, the search for the excitation origin during the ablation procedure can be accelerated by giving multiple ranked estimates.

The initial results reported here are not yet on a par with the accuracy that can be achieved with approaches incorporating the patient-specific geometry. This might be due to several reasons that will be discussed in the following and provide potential for future improvement. 

We introduced ScaleNet as a method for detecting the start and end of the excitation propagation in the heart. Usually, classic signal processing methods are utilized to detect these points from the surface ECG. Nevertheless, the boundaries of the QRS complex do not necessarily correspond to the actual intracardiac activation time extrema. Having the ground truth from simulations, it seemed obvious to use this information for the training of an additional CNN to overcome exactly this discrepancy. We could show that, on the simulated data set, ScaleNet performed with high accuracy. 

For localization, both, RegNet and ClassNet yielded low errors (median localization error below 2.5\,mm), which appears noteworthy considering the amount of geometrical variations and the dependency on ScaleNet's performance. Existing works achieve (mean) errors of around 4-15\,mm considering less geometrical variation~\cite{Arrieula_2021,Yang-2018-ID12792,unknown-0000-ID16466}, where CNNs achieve better results than other classical machine learning methods, e.g., Random Forests, Support Vector Machines, etc. Thus, literature is in accordance with the finding that our CNNs can in general yield a precise localization without patient-specific geometrical information (i.e., without tomographic imaging). Furthermore, our CNNs seem to generalize enough to cope with signals from new and unseen geometries.

We were surprised that our methods yielded usable results on clinical data although trained only on simulated data. We originally expected that transfer learning techniques would need to be applied. In a work published after the completion of out study, the authors also found simulated data to be at least beneficial for the performance on clinical data~\cite{Doste_2022}. If clinical data was available in a larger quantity, as it was to the authors of~\cite{Doste_2022}, results could have further improved by using a part of the clinical data as training data, incorporated into the training process. Unfortunately at the current state, localization errors obtained from our implemented methods could be as high as 110\,mm on clinical data. Comparing our found median localization errors of 32.6\,mm/40.3\,mm (SVD-methods) on clinical data with existing works based on learning methods, we find methods with lower errors of 10.9\,mm in~\cite{Gyawali-2020-ID14304} (considering only the LV), down to 10.3\,mm in~\cite{Yang-2018-ID12792} (but using patient-specific geometries), and with similar errors of around 30\,mm in~\cite{Alawad-2019-ID12834} (again using patient-specific geometries). In related work~\cite{Schuler2022_1000141624}, we calculated the performance on different ECG imaging techniques, also using the patient-specific geometry. Median errors were between 12.8\,mm and 28.8\,mm, which is also better than our approach but at the cost of needing a geometrical information. Although ClassNet's median top-3 error drops down to approximately 20\,mm, especially the results in~\cite{Gyawali-2020-ID14304} which yield an approximately 10\,mm respectively 20\,mm lower error than ours underline the need for further investigations and optimizations of our methods\footnote{Although we expected higher errors in our work than in~\cite{Gyawali-2020-ID14304} (as they only consider the LV endocardial wall and calculate Euclidean and not geodesic errors), we did not expect such a huge gap between the errors.}. It seems that the domain gap between our simulations and the clinical data is not entirely closed. We could not precisely pinpoint why the CNNs failed to adequately localize the excitation origin in our clinical data set. We do not consider major design errors to be responsible for this since the performance on simulated data was good. However, fine-tuning could improve results. Until now we did not incorporate any adaptions from our first experiment with the data set apart from the SVD-filtering step. Optimized network structures, training, hyperparameters, BSPM processing steps are only examples for potential improvement steps helping to close the domain gap completely. Furthermore, a lead reduction might deliver further improvements~\cite{Lai_2021}. For ClassNet, another choice for the number of triangles might be optimal for this learning task. Results could improve if the number of classes (and thus the dimension of the optimization problem) was changed and if we would apply transfer learning on only a few clinical examples that are unfortunately not available.

The calculated the top-3 error of ClassNet should be interpreted with care. Although interpreting the output of ClassNet as multiple solutions ranked by their likelihood decreased the localization error of the clinical data set, this stays an empirical finding. CNNs are known to be over-confident and not well calibrated~\cite{Guo2017} and thus, their likelihood output can be untrustworthy~\cite{Gal2016}. However, most classification tasks work with one-hot-encoded labels opposed to our barycentric coordinates where the ground truth has more than one value larger than zero. This might lead to a different learning behaviour of ClassNet and therefore might also allow for a different interpretation of it's output. Uncertainty estimation of ClassNet's output should definitely be a focus of future work in order to understand the trustworthiness of the multiple solutions in detail.

Apart from these, there are other possible reasons: first, idealizations and missing details in the excitation propagation model, in the forward model and in the heart geometries might cause the CNNs to learn undesired behavior. In this study, approximately 1.8 million BSP images were generated to train different CNNs. Due to limited time and hardware resources, this required using less detailed modeling techniques: on the one hand, a fast excitation propagation simulation method; on the other hand, a homogeneous torso model for the forward calculation. Both choices limit the fidelity of the simulation results~\cite{neic17,Rudy-2015-ID12049}. Nevertheless, these approaches enabled the generation of this high number of training samples required for deep learning approaches under the given time and hardware resources. Moreover, the extent of details needed for the training cannot be estimated a priori. The CNN learning undesired behaviour is at least partially substantiated by the improved results after SVD filtering which seems to leave out unnecessary details and should be addressed by further regularization strategies in the future. Additionally, we cannot exclude problems originating from the ventricular shape model covering only parts of the real-world variability and the right ventricular wall thickness being homogeneously (2\,mm). Another possible reason for the high error values lies in our clinical data itself. Unfortunately, the signals were pre-cropped by clinicians. Possibly, they excluded too much of the QRS complex and this leads then to missing but important parts of the signal. Actually, ScaleNet detected negative time points for the excitation start in most of the cases, which could indicate a wrong manual cropping. In all samples of the simulated data set, ScaleNet detected the excitation start only once before the actual start from the simulations with an error of 0.6\,ms. Lastly, we cannot be sure about ScaleNet's performance on clinical data as we had no invasive measurements of the actual activation start and end available. Additional studies containing a comparison between ScaleNet and other methods -- also using the QRS information -- could help to learn about the sensitivity of the localization methods on the results from the start and end of excitation propagation in the heart. A third possible reason for the impaired results on clinical data is that the patients suffered from diseases that were not covered with our simulations. These could for example be slow conducting areas or pathological anatomical variations (only partly covered by the underlying shape model) that were not included in our large simulated data set. These missing training data reflect a potential undercomplexity of our simulations and could therefore lead to a knowledge gap for the CNNs after training. Finally, it should be noted that the excitations in the clinical data were induced by a pace maker and not by an ectopic region. Up to now, we do not know how transferable our method is to patients with self-triggering ectopic regions.


\section{Conclusion}
\label{conc}
We can conclude that CNNs are an appropriate method for the bi-ventricular localization of excitation origins using BSP images. The Cobiveco ventricular coordinates are a useful tool to visualize results on a generic or an existing patient-specific geometry as well as helping to define and translate the segments (and in our case the fuzzy classes) between different patients. We addressed the need of large data sets for deep learning techniques by an optimized simulation pipeline generating 1.8 million data points. The domain gap between simulations and clinical data was small but could not entirely be closed by our selection of simulation and processing methods. However, the methods worked surprisingly well on unseen clinical data.

With further adjustments of the method (introduction of further pathologies into the simulations, a more sophisticated cropping of patient data, the use of transfer learning, in-depth uncertainty handling), our approach has a high potential for being able to translate the CNNs only trained on simulated data to a clinical application. Our approaches have the potential for being used as a fast pre-screening method accelerating the clinical procedures as our fuzzy classification approach offers several ranked solutions which are likely useful during the intervention. This can give the cardiologists at least a hint where to look for the excitation origin, which is today only possible to a certain extent. This can accelerate ablation procedures, save costs, increase throughput while at the same time improving the treatment quality for the patient.

\section*{Acknowledgement}
Steffen Schuler's work was supported by the German Research Foundation [grant DO 637/21-1]. We thank for support by the NVIDIA Academic Hardware Grant.

\printcredits

\bibliographystyle{cas-model2-names}
\bibliography{refs}

\end{document}